\documentclass[creativecommons,noderivs]{eptcs}
\providecommand{\event}{ML 2015} 

\usepackage{url}
\usepackage{hyphenat}
\usepackage{mdwtab}
\usepackage{mdwlist}
\usepackage{amssymb}
\usepackage{comment}
\usepackage{prooftree}

\usepackage{listings}
\usepackage{breakurl}

\lstdefinestyle{sOcaml}{language=[Objective]Caml,
  literate={+}{{$+$}}1 {/}{{$/$}}1 { * }{{$*$}}1
           {=}{{$=\ $}}1
           {>}{{$>$}}1 {<}{{$<$}}1
           {<>}{$\not=$}1
           {->}{{$\rightarrow$}}2 {>=}{{$\geq$}}2 {<-}{{$\leftarrow$}}2
           {<=}{{$\leq$}}2
           {==}{{==}}2
           {==>}{{$\mapsto$}}2
           {|}{{$\mid$}}1
           {ALT}{{$\mid$}}2
           {'a}{$\alpha$}1
           {+'a}{$\textrm{+}\alpha$}1
           {'b}{$\beta$}1
           {+'b}{$\textrm{+}\beta$}1
           {'w}{$\omega$}1
           {forall}{$\forall$}1
           {...}{\ldots}1
           {TRB}{\mbox{\ensuremath\lceil}}1
           {TRE}{\unskip\mbox{\ensuremath\rceil}}1
           {SRB}{\mbox{\ensuremath\lfloor}}1
           {SRE}{\unskip\mbox{\ensuremath\rfloor}}1
           {\#\#\#}{{$\leadsto$}}3
}
\lstnewenvironment{code}{\lstset{basicstyle={\sffamily\footnotesize}}}{}
\makeatletter
\newcounter{mylstnumber}
\def\mynumberreset{\c@mylstnumber=0}
\def\mylabel#1{\expandafter\edef\csname mr@#1\endcsname{(\the\value{mylstnumber})}}
\def\myref#1{\csname mr@#1\endcsname}
\def\mynonumbering{\phantom{(\the\value{mylstnumber})}\kern 10pt}
\lst@AddToHook{Init}
 {\def\mynumbering{\global\advance \c@mylstnumber 1\relax
(\the\value{mylstnumber})\kern 10pt}}
\lst@AddToHook{EveryPar}
 {\mynumbering\let\mynumbering\mynonumbering}
\makeatother
\lstMakeShortInline[columns=fullflexible]|

\newtheorem{proposition}{Proposition} 

\newcommand{\aside}[1]{\ignorespaces}

\newcommand\entail{\vdash}

\newcommand{\Ju}[4]{#1 \entail^{#2}\; \textsf{#3} \::\: \textsf{#4}}
\newcommand{\Bra}[1]{\texttt{<}#1\texttt{>}}

\begin{document}

\title{Generating Code with Polymorphic let\\
  A Ballad of Value Restriction, Copying and Sharing}
\author{Oleg Kiselyov
\institute{Tohoku University, Japan}
\email{oleg@okmij.org}
}
\def\authorrunning{Oleg Kiselyov}
\def\titlerunning{Generating Code with Polymorphic let}
\maketitle

\begin{abstract}
Getting polymorphism and effects such as mutation to live together in the
same language is a tale worth telling, under the recurring refrain of
copying vs.~sharing. We add new stanzas to the tale, about the ordeal
to generate code with polymorphism and effects, and be sure it
type-checks. Generating well-typed--by--construction polymorphic
let-expressions is impossible in the Hindley-Milner type
system: even the author believed that.

The polymorphic-let generator turns out to exist. We present its
derivation and the application for the lightweight implementation
of quotation via a novel and unexpectedly simple source-to-source
transformation to code-generating combinators.

However, generating let-expressions with polymorphic functions demands
more than even the relaxed value restriction can deliver. We need a
new deal for let-polymorphism in ML. We conjecture the weaker
restriction and implement it in a practically-useful code-generation
library. Its formal justification is formulated as the research program.
\end{abstract}

\section{Introduction}

This paper revolves around code generation, namely, generating typed,
higher-order code for languages such as OCaml. Specifically we deal
with one approach to code generation: staging (recalled in
\S\ref{s:staging}), and the lightweight way of implementing it via
code-generating combinators. In our approach, the generated code is
assured to be well-formed and well-typed by construction: attempts to
produce ill-typed fragments are reported when type-checking the
generator itself. In contrast, the post-validation used, for example,
in Template Haskell \cite{sheard-template}, type-checks the code only
after it has been completely generated. The errors are thus reported
in terms of the generated code rather than the generator, breaking
the generator's abstractions\footnote{\relax
Post-validation is hence similar to run-time failure of ill-typed
code in dynamically-typed languages. However, with a run-time error 
we can get a stack trace, etc. On the other hand, when post-validating
the (typically large and obfuscated) generated code,
the generator is long gone and its state can no longer be examined.}.

However, staging here is the lens through which to look at the old
problem of let-generalization. The unexpected interactions of
polymorphism and staging brings into focus the `too obvious' and hence
rarely mentioned assumptions of the value restriction. Generating
code that contains polymorphic let-expressions is a non-contrived,
real-life application that requires let-generalization of effectful
expressions~-- going beyond what even the relaxed value
restriction offers.  Staging thus motivates further work on
the seemingly closed topic of let-generalization in the presence of
effects.

Although program generation is a vast area, surprisingly there has
been very little research on typed-assured code generation with
polymorphic let.  To our knowledge, \cite{kobayashi-poly} is the first
paper that brings up a staged calculus that has both
polymorphism and mutable cells. It is motivated by the unexpected
interaction of polymorphism and staging that we describe in
\S\ref{s:poly-stage}. There are many systems for typed code generation
(see \cite[\S5]{hygienic-combinators} for the recent overview) yet
polymorphic-let expressions are not included in the target language.
The only related, albeit quite remotely, is the work
\cite{self-interp-f-omega} on typed self-interpreters, which does include the
representation of polymorphic expressions as code~-- but lacks any
effects. That work is based
on System F$_\omega$, which is difficult to use in practice, in part
because it lacks type inference. In contrast, in our code generation
approach all types are inferred.

\paragraph{Contributions} First, the paper presents a new translation
from the staged code~-- with quotations, unquotations and cross-stage
persistence~-- to quotation-free expressions over code-generating
combinators. The translation is remarkably simpler than the other
unstaging translations. It also translates quoted let-expressions to
let-expressions, for the first time giving the chance to
generate polymorphic let-expressions, well-typed by
construction. Second, we present the first library of typed code
combinators whose target language includes polymorphic let-bindings
and effects. The library requires no first-class polymorphism, no type
annotations and, combined with the unstaging translation, is
suitable for implementing staging by source-to-source translation to
combinators.  The library solves the problem that the author claimed
in 2013 to be unsolvable \cite{ml2013-talk}.

Although the translation and the library are already practically
useful, their formalization requires deeper understanding of
polymorphism and effects. The paper proposes a research program, which
will have to open the old value-restriction wounds and could finally
heal them. Thus we end up posing more questions~-- the questions that
could not have been asked before.

The paper starts with extensive background. \S\ref{s:poly} recalls
let-polymorphism and the ways to restrict it in the presence of
effects. That section describes the copying-sharing dichotomy that
reverberates through the rest of the paper. \S\ref{s:staging}
introduces staging, using MetaOCaml as an example, and
\S\ref{s:poly-stage} describes the unexpected interactions of staging
and polymorphism. We then describe in \S\ref{s:translation} the novel
translation that systematically eliminates quoted expressions,
replacing them with applications of code combinators. Alas, the
translation does not seem work for polymorphic let-expression, as shown in
\S\ref{s:problem}. It can be made to work; \S\ref{s:new-translation}
explains how. As in the ordinary ML, reference cells and polymorphism
is a dangerous mix; our translation hence needs some sort of a 
restriction, weaker than even the relaxed value restriction.
\S\ref{s:new-level-restriction} discusses the solutions and the many
follow-up problems.

We will be using OCaml throughout the paper for concreteness. However
the discussion equally applies to any other typed, higher-order
language with polymorphism and effects.

The complete code accompanying the paper is available at

\url{http://okmij.org/ftp/meta-programming/polylet.ml}

\section{Background}
\label{s:background}
This background section recalls let-generalization; its problems in
the presence of effects; staging; and the unexpected interaction of
generalization and staging that calls up the assumptions of the value
restriction. The section introduces the running examples used later in
the paper.

\subsection{Let-polymorphism}
\label{s:poly}

Since the early days of LISP and ISWIM \cite{landin-700},
let-expressions let us introduce and name common or notable expressions
which are used, often repetitively, later in the code. Here is a simple example:
\begin{code}
let x = [1] in
(2::x,3::x)
\end{code}
It may be regarded as a sort of a `macro' that expands into
\begin{code}
(2::[1],3::[1])
\end{code}
In fact, such a `macro-expansion'~-- copying (inlining) the let-bound expression
into the places marked by the let-bound variable~-- is the meaning
given to let-expressions in Landin's ISWIM \cite{landin-700}.  The
alternative to this copying, or substitution-based semantics is
sharing. It views (1) as
introducing the expression |[1]| that is shared across all
occurrences of |x|. Hence the two lists in (2)
share the common tail. Copying vs. sharing reverberates throughout the
paper as the constant refrain. If our program simply prints out (2),
the two semantics are indistinguishable. The equivalence lets the
compiler choose inlining or sharing as fits.

Likewise, the code
\begin{code}
let x = [] in
(2::x,"3"::x)
\end{code}
may be viewed as a macro that expands into
\begin{code}
(2::[],"3"::[])
\end{code}
It is tempting to also regard (3) as the sharing of the empty list
across the two components of the returned pair. Unlike (2), however,
|[]| in (4) has different types: namely, |int list| in the
first component vs.  |string list| in the second. Thus comes the
problem of what type to give to the shared value and to |x|.

The answer developed by  Milner \cite{milner-theory} was polymorphism:
the  same  expression  that  occurs  in~--  has  been  copied  into~--
differently typed  contexts may  be shared and  given the  common, the
most general,  polymorphic type (see  also the extended  discussion in
\cite{cardelli-basic}).  The  empty list |[]|  has the type  |'a list|
to be fully determined by the context;
|'a| is the placeholder: a  (unique) type variable.
In (4), the contexts determine the type as
|int list| and |string list|, respectively. In (3), the context of the
right-hand side (RHS) of the let-binding has not determined what |'a list|
should  be.  In that  case,  the  type  is \emph{generalized}  to  the
polymorphic type schema: |forall'a. 'a list|.

Formally, the typing of let-expressions is represented by the
(GenLet) rule below. The rule is written in terms of the judgments
$\Gamma \entail \textsf e: \textsf t$ that an
expression |e| has the type |t| in the type environment $\Gamma$
(which lists the free variables of |e| and their types).

\medskip
\begin{prooftree}
\Gamma \entail \textsf{e} : \textsf{t}
\qquad
\Gamma, \textsf{x}:GEN(\Gamma,\textsf{t}) \entail \textsf{e'} : \textsf{t'}
\justifies
\Gamma \entail \textsf{{\bf let} x = e in e'} : \textsf{t'}
\using GenLet
\end{prooftree}
~~~~
\begin{prooftree}
\textsf x : \forall \alpha_1\ldots\alpha_n.\textsf t \in \Gamma
\justifies
\Gamma \entail \textsf x : t\{\alpha_1=\textsf t_1\ldots
   \alpha_n=\textsf t_n\}
\using Inst
\end{prooftree}

\medskip
\noindent
The generalization function $GEN(\Gamma,\textsf t)$ for the type |t|
with respect to the type environment $\Gamma$ quantifies the free type
variables of |t| that do not occur as free in $\Gamma$:
\[
GEN(\Gamma,\textsf t) = \forall \alpha_1\ldots\alpha_n.\textsf t
\qquad \textrm{where} \{\alpha_1\ldots\alpha_n\} = FV(\textsf t) - FV(\Gamma)
\]
where $FV(\cdot)$ denotes the set of free variables. When a variable
with the polymorphic type schema such as |x: forall 'a. 'a list| in
(3) is used in an expression, e.g., |3::x|, the schema is converted to
a more specific type, |int list| in our example: see the rule (Inst).
The underlying assumption is that the value named by |x| indeed has
the same representation for all instances of the polymorphic type
schema and hence may be shared, even across differently-typed
contexts; the instantiation is a purely type-checking-time operation
that behaves like identity at run-time.  One may say that the
motivation of polymorphism is to extend the equivalence of the copying
and sharing semantics to the cases like (3).

Side-effects break the equivalence of copying and sharing.
\begin{code}
let x = begin printf "bound"; [1] end in
(2::x,3::x)
\end{code}
If the right-hand side is copied, substituting for the occurrences of
|x| in (5), the string |"bound"| is printed twice.  If the RHS is
first reduced, however, (5) turns to the earlier (1), where |x| is
bound to the value that can be either shared or copied. Hence the
copying/sharing equivalence holds even in the case of (5), if we
regard variables as bound to values~-- as we do in call-by-value
languages\footnote{\relax
Another way to restore the equivalence is to regard \textsf{x} as
bound to an expression that is evaluated only at the places of
\textsf{x}'s occurrence. That was the idea of Leroy's call-by-name
polymorphism \cite{Leroy-cbn-poly}.}. The polymorphic case 
should likewise be unproblematic:
\begin{code}
let x = begin printf "bound"; [] end in
(2::x,"3"::x)
\end{code}

The polymorphic equality of OCaml can distinguish sharing and copying:
\begin{code}
let x = [1] in
x == x 
\end{code}
whereas (7) returns the result |true|, the expression |[1] == [1]|
produces |false|. Another, universal way to distinguish sharing and
copying uses mutable data structures, in particular,
mutable cells \cite{baker-equal}. Let's define
\begin{code}
let rset : 'a list ref -> 'a -> 'a list = fun r v ->
  let vs' = v :: !r in
  r := vs'; 
  vs'
\end{code}
that prepends the value |v| to the list stored in the reference cell |r|,
stores the new list in the cell and returns it. Then
\begin{code}
let x = ref [1] in
(rset x 2, rset x 3)
### ([2; 3; 1], [3; 1])
(rset (ref [1]) 2, rset (ref [1]) 3)
### ([2; 1], [3; 1])
\end{code}
produce the different results as shown underneath the
expressions. Since the distinction between copying and sharing is
generally visible, there is no longer freedom of choosing between the
two. OCaml uses sharing for |let|-expressions, performing inlining
(copying) only when it can see the equivalence.

The example paralleling (3) however does not type-check.
\begin{code}
let x = ref [] in
(rset x 2, rset x "3")
(* Does not type-check! *)
\end{code}
\mylabel{e:refpoly0}
As we have just seen, with reference cells, sharing and copying differ
and the OCaml compiler has to use the default sharing. Had
the expression type-checked, at run-time |rset x "3"| would modify
the empty list stored in |x| by prepending the string |"3"| to it. The
expression |rset x 2| will then try to prepend the integer |2| to the
contents of |x|, which by that time is the string list
|["3"]|. Clearly that is a program that has ``gone wrong''. We should
well remember this example: we shall be seeing it, in different guises,
all throughout the paper.

Although the RHS of the let-binding in (10) has the type |'a list ref|
with the type variable that could be generalized, it should not be, to
prevent (10) from type-checking.  Intuitively, sharing and copying of
a reference cell have different semantics, hence it should not get the
polymorphic type schema.

The danger of giving reference cells a polymorphic type has been
recognized early on \cite{tofte-thesis}. So has the problem of how to
restrict (GenLet) from applying to ``dangerous''
expressions. The most straightforward solution, used in the early ML
and OCaml for a long time, was to limit reference cells to base
types only. The restriction made it impossible however to write any
polymorphic function that internally uses reference cells.  A good
overview of less draconian approaches is given in
\cite{garrigue-relaxing}. The most widely implemented, because of its
balance of expressiveness with sheer simplicity, is the value restriction
\cite{wright-simple}: applying (GenLet) only to those
let-expressions whose RHS is syntactically a value. Since |ref []| in
(10) is not a value, |x| is not generalized and its occurrences in
differently typed contexts will raise the type error. On the other
hand, |[]| in (3) is a value and |x| there does get the polymorphic
type. Strictly speaking, |x| in (6) should not be generalized
either. However, it is syntactically obvious that the printing effect
has no contribution to the result of the containing expression. The RHS of (6)
is what is called `non-expansive'. OCaml generalizes
non-expansive expressions, not just values.

Although the value restriction on balance proved expressive enough for
many ML programs, as OCaml gained features such as objects, polymorphic
variants and a form of first-class polymorphism (enough to support
polymorphic recursion), the restrictiveness of the value restriction
was felt more and more acutely. Against this backdrop, Garrigue introduced the
`relaxed value restriction' \cite{garrigue-relaxing}, which we briefly
overview below as we will be relying on it.

The relaxed value restriction explores the close analogy between 
type instantiation and subtyping. It can also be justified from the
point of view of copying-sharing: a value occurring in
differently-typed contexts may be let-bound and shared if it can be
given the `common type', the supertype of the types expected by the
contexts. The coercion to a subtype, like the type instantiation, is
a compile-time--only operation, behaving as identity at run-time. Suppose
a value has the type |zero c| where |zero| is the empty type, and it
can be \emph{coerced} by subtyping to the type |t c| for any |t|.
We may as well then give the
variable that is let-bound to the value the type |forall'a. 'a c|, which can
then be \emph{instantiated} to |t c|. Since |zero| is (vacuously) coercible
to any type, a value of the type |zero c| can be coerced to |t c| only
when the type |zero| occurs covariantly in |zero c|. Hence the relaxed
value restriction: If the expression |e| in |let x = e in e'| has a
type with covariant type variables (which do not occur in the typing
context), they are generalized in the type inferred for |x|. (The
actual implementation is somewhat more restrictive: see
\cite{garrigue-relaxing} for details.)

For example, |x| below is generalized
\begin{code}
let x = let r = ref [] in !r in
(2::x,"3"::x)
\end{code}
despite the fact the RHS is an expression -- moreover, the expression
whose result comes right from a reference cell. Still, the result has the
type |'a list| whose type variable is covariant (with |List.map| being
the witness of it). On the other hand, the type of reference cells |'a ref| is
non-variant and hence |x| in (10) remains ungeneralized. The relaxed
value restriction applies not only to built-in data types but also
to user-defined and abstract ones:
\begin{code}
type +'a mylist = List of 'a list
let mklist : 'a list -> 'a mylist = fun x -> List x
let mycons : 'a -> 'a mylist -> 'a mylist = 
   fun x -> function List l -> mklist (x::l)
let x = mklist [] in
(mycons 2 x, mycons "3" x)
\end{code}
Although the RHS of the let-binding is an expression, |x| is
generalized because the type |'a mylist| is covariant in |'a|.  It
is declared to be covariant, by the |+'a| covariance
annotation. The compiler will check that the RHS of the type
declaration really uses the type variable |'a| covariantly. The
compiler can also infer the variance, hence the annotations are
normally omitted. They are necessary only for abstract types, whose
declaration lacks the RHS.

Overall, the relaxed value restriction turned up even better balanced,
accommodating not just polymorphic functions but polymorphic data
(including row-polymorphic data such as extensible records and
variants), whose construction often involves computations. The relaxed
value restriction was almost enough for implementing staging via code
combinators~-- but not quite, as we see in
\S\ref{s:new-level-restriction}. Let us first review staging.

\subsection{Staging}
\label{s:staging}\mynumberreset

Staging is an approach to writing programs that generate programs;
it may be regarded as a principled version of Lisp quotation. 
For example, MetaOCaml lets us quote any OCaml
expression, by enclosing it within |.<| and |>.| brackets:
\begin{code}
let c = .<1 + 2>.
### val c : int code = .<1 + 2>. 
\end{code}
A quoted expression, representing the generated code, is the value of
the type |'a code|, and can be bound, passed around, printed~-- as well
as saved to a file and afterwards compiled or evaluated. For that reason, the
code value such as |c| is called a `future-stage' expression (or,
an expression at level 1), to contrast to the code that is being
evaluated now, at the present, or 0, stage. An expression that
evaluates to a code value can be spliced-in (or, unquoted, in Lisp
terminology) into a bigger quotation:
\begin{code}
let cb = .<fun x -> .~c + x>.
### val cb : (int -> int) code = .<fun x_1  -> (1 + 2) + x_1>. 
\end{code}
The spliced-in expression is marked with |.~|, which is called an escape. The
generated code can be executed by the function |run| (in the module
|Runcode|), reminiscent of Lisp's |eval|:
\begin{code}
open Runcode
val run :: 'a code -> 'a

let cbr = run cb
### val cbr : int -> int = <fun>
cbr 2
### - : int = 5
\end{code}
Running |cb| hence compiled the |cb| code of a function into an
|int->int| function that can be invoked at the present level. As one
expects, running the code indeed invokes the compiler and the dynamic
linker. The |run| operation hence lets us generate code at run-time and
then use it~-- in other words, it offers run-time code specialization.

When generating functions it is natural to require that the behavior
of the resulting program should not depend on the choice of names
for bound variables.  For example,
\begin{code}
let c1 = .<fun x -> .~(let body = .<x>. in .<fun x -> .~body>.)>.
### val c1 : ('a -> 'b -> 'a) code = .<fun x_1  -> fun x_2  -> x_1>. 

let c2 = .<fun y -> .~(let body = .<y>. in .<fun x -> .~body>.)>.
### val c2 : ('a -> 'b -> 'a) code = .<fun y_3  -> fun x_4  -> y_3>. 
\end{code}
The expressions |c1| and |c2| should build the code that behaves the
same when evaluated. This is indeed the case, as one can see from the
generated code, printed underneath.  If we write this example with
quotations in Lisp, the expressions are no longer equivalent:
whereas |c2| generates the code for the K combinator, |c1| builds
a function that takes two arguments returning the second one.
Lisp quotations are hence not hygienic.

When generating code for a typed language, it is also natural to require
that the produced code is type-correct by construction. For that
reason, the code type is parametrized by the type of the generated expression,
as we saw for |c|, |cb|, etc. The formal treatment of
type soundness is well covered in
\cite{taha-environment,calcagno-ml-like} and will be briefly reminded
of in \S\ref{s:poly-stage}.

MetaOCaml has yet another facility, which has no special syntax and is
easy to overlook. Let us look again at |.<1 + 2>.| and ponder the
addition operation there. In the ordinary OCaml expression |1+2|, the
addition is the ordinary function, defined in the |Pervasives| module.
The addition in |.<1+2>.| refers to exactly the same function:
MetaOCaml permits any value of the generator to appear in the
generated code. This is called ``cross-stage persistence'' (CSP) (see
\cite{taha-environment} for more discussion).  One may think of CSP
identifiers as references to `external libraries'.

The trivial code for the addition of two numbers has already
demonstrated how wide-spread CSP is. Let us show a more explicit
example of CSP, brought about by the function
\begin{code}
let lift : 'a -> 'a code = fun x -> .<x>.
\end{code}
The following example then produces the code as shown (compare with (2)):
\begin{code}
.<fun x -> .~(lift (1 + 2)) + x>.
###  - : (int -> int) code = .<fun x_1  -> (* CSP x *) Obj.magic 3 + x_1>. 
\end{code}
In contrast to (2), here the addition of |(1+2)| is done at the code
generation time; the generated code includes the computed value. 
CSP hence lets us do some of the future-stage computations at the
present stage, and hence generate more efficient code. The bizarre
|Obj.magic| appearing in the generated code is the artifact of
printing. The following code
\begin{code}
.<fun x -> .~(let y = 1 + 2 in .<y>.) + x>.
### - : (int -> int) code = .<fun x_2 -> 3 + x_2>. 
\end{code}
(where the CSP identifier |y| is known to be of the |int| type) produces the
more obvious result. 

Our refrain of copying vs. sharing repeats for CSP. When a value from
the present stage is used at a future stage, do the two stages share
the value or does the future stage get a separate copy? Unfortunately this
issue is not discussed in the literature let alone formally
addressed\footnote{the exception being the work \cite{kobayashi-poly} which was
  inspired by the problem we discuss in \S\ref{s:poly-stage}.}~--
which is a pity since it is responsible for the unexpected soundness
problem to be described in the next section. The case of a global
(library) identifier seems clear: code such as |.<succ 3>.|
contains the identifier |succ| that refers to the same library
function it does at the present level. Whether that function is shared
or copied between the present-stage and the generated code depends on
the inlining strategy of the compiler and the static vs. dynamic
linking. One could expect sharing/copying to be equivalent in this case.

The cross-stage persistence of a locally-created value is much less
clear\footnote{The right-associative infix operator \textsf{@@} of low
precedence is application: \textsf{f @@ x + 1} is the same as
\textsf{f (x + 1)} but avoids the parentheses. The operator is the
analogue of \textsf{\char`$} in Haskell.}: 
\begin{code}
let cs = 
  let z = string_of_float @@ Sys.time () in
  .<print_endline z>.
### val cs : unit code = .<Pervasives.print_endline "0.051">. 
\end{code}
One may imagine that the code value |.<print_endline z>.|
(represented, say, as an AST) contains the pointer to a string
allocated in the running-program heap~-- the same pointer that is
denoted by the local variable |z|. Then |run cs|
will print the value of that string on the heap.  The present and the
future stage hence share the string. Rather than running |cs| however, 
we may save it to a file, as library code for use
in other programs. In this case, when the generated code is evaluated
the generator program is long gone, along with its heap. Therefore, when
storing a code value to a file we must serialize all its CSP values,
creating copies. In the upshot, cross-staged persistent library identifiers are
always shared; other CSP values are shared if the code value is |run|,
and copied otherwise. The semantics of CSP is indeed intricate.
We have just described the CSP implementation in the extant MetaOCaml;
there is an ongoing discussion of it and its possible improvements
\footnote{\url{http://okmij.org/ftp/ML/MetaOCaml.html\#CSP}}.

The question of sharing vs. copying CSP becomes non-trivial when
the CSP value is mutable:
\begin{code}
 let r = ref 0 in
 let cr = .<incr r>. in
 run cr; run cr; !r
### - : int = 2
\end{code}
\mylabel{e:rcsp}
Mutable CSP values naturally arise when run-time specializing 
imperative code. They can be used for cross-stage
communication, e.g., counting how many times the code is run, 
as shown in \myref{e:rcsp}~--
which works as intended only with the shared CSP.
Sharing of mutable CSP values is also responsible for the unexpected problem
with let-polymorphism, detailed next.

\subsection{Let-polymorphism and Staging}
\label{s:poly-stage}
\mynumberreset

For a long time let-polymorphism and staging were considered orthogonal
features. It is not until 2009 that their surprising interaction was
discovered\footnote{\relax
  \url{http://okmij.org/ftp/meta-programming/calculi.html\#staged-poly}};
it has not been formally published. Before describing this
interaction, we first briefly remind the type system of a staged
language, on a representative subset of MetaOCaml.

Staging adds to the base language the expression forms for brackets
\textsf{\Bra e} and escapes \textsf{\~{} e} and the type of code
values |t code|. We use the meta-variable |x| for variables, |e| for
expressions, |v| for values, and |t| for types. The type system 
is essentially the standard, Figure \ref{f:type-system}. It is derived
from the type system of \cite{taha-environment} by replacing the
sequence of no-longer used classifiers with the single number, the
stage level. (Since brackets may nest, there may be
an arbitrary number of future stages.) The judgments have
the form $\Ju{\Gamma}{n}{e}{t}$: they are now indexed by the level of
the expression; the level is incremented when type-checking the
expression within brackets and decremented for escapes. The identifiers
within the typing environment $\Gamma$ are now indexed by the level at
which they are bound. The (GenLet) rule reflects the value restriction.

\begin{figure}
\centering
\begin{prooftree}
\Ju{\Gamma}{n}{e}{t'$\to$t} \quad
\Ju{\Gamma}{n}{e'}{t'}
\justifies
\Ju{\Gamma}{n}{e\ e'}{t}
\end{prooftree}
~~~~
\begin{prooftree}
\Ju{\Gamma, \textsf x ^n:\textsf{t'}}{n}{e}{t}
\justifies
\Ju{\Gamma}{n}{{\bf fun} x $\to$ e}{t'$\to$t}
\end{prooftree}
\\[3\jot]
\begin{prooftree}
\Ju{\Gamma}{n}{v}{t}
\qquad
\Ju{\Gamma, \textsf x ^n:GEN(\Gamma,\textsf{t})}{n}{e'}{t'}
\justifies
\Ju{\Gamma}{n}{{\bf let} x = v in e'}{t'}
\using GenLet
\end{prooftree}
\\[3\jot]
\begin{prooftree}
\Ju{\Gamma}{n+1}{e}{t}
\justifies
\Ju{\Gamma}{n}{\Bra e}{t code}
\using Bracket
\end{prooftree}
~~~~
\begin{prooftree}
\Ju{\Gamma}{n}{e}{t code}
\justifies
\Ju{\Gamma}{n+1}{\~{} e}{t}
\using Escape
\end{prooftree}
~~~~
\begin{prooftree}
\Ju{\Gamma}{n}{x}{t}
\justifies
\Ju{\Gamma}{n+1}{x}{t}
\using CSP
\end{prooftree}
\caption{Type system of a staged language}
\label{f:type-system}
\end{figure}

Staging thus contributes the three rules (Bracket), (Escape) and (CSP)
and the indexing of the environment and the judgments by the stage level.
If the program has no brackets, the stage level stays at 0 and the type
system degenerates to the one for the (subset of the) ordinary
OCaml. Moreover, except for the three staging-specific rules, the
rest are the ordinary OCaml typing rules, uniformly indexed
by the stage level. Thus, aside from brackets, escapes and CSP, the
type-checking of the staged code proceeds identically to that for the
ordinary code. In particular, let-expressions within brackets are
handled and generalized the same way they do outside brackets. For
example:
\begin{code}
.<let x = [] in (2::x,"3"::x)>.
.<let f = fun x -> x in (f 2, f "3")>.
\end{code}
\begin{code}
.<let x = ref [] in (rset x 2, rset x "3")>.  (* Does not type-check! *)
\end{code}
It appears hence that let-generalization and staging are orthogonal
features.

Consider however the following code
\begin{code}
.<let f = fun () -> ref [] in 
     (rset (f ()) 2, rset (f ()) "3")
>.
\end{code}
The type-checker accepts it and infers the type 
|(int list * string list) code|. The variable |f| hence gets the
polymorphic type. After all, the RHS of the let-binding is syntactically
the (functional) value. There is really nothing wrong with (3):
|f| can be either copied or shared across its uses without the change
in semantics: the invocation |f ()| in either case will produce a
fresh reference cell holding an empty list, later
modified by prepending either |2| or |"3"| to its contents.

Now consider the simple modification, along the lines of (6) in
\S\ref{s:staging}:
\begin{code}
let cbad =
.<let f = fun () -> .~(lift (ref [])) in 
     (rset (f ()) 2, rset (f ()) "3")
>.
run cbad
### Segmentation fault
\end{code}
It is also accepted, with the same inferred type~-- for any version of
MetaOCaml including the current one. The RHS of the
let-binding is still syntactically a function; we merely modified its
body. Running that code however ends in the segmentation fault. One
should not be surprised: we have managed to generate and type-check
\myref{e:refpoly0} from \S\ref{s:poly}, the canonical example of the
unsoundness of polymorphism for reference cells.

Thus staging breaks the restriction of the value restriction,
unleashing the unsound generalization. If we re-examine the value
restriction we now notice an assumption, which is rarely stated
explicitly: there are no literals of reference types; every expression
of the type |t ref| is not syntactically a value. Cross-stage
persistence, however, lets one stage share its values with a future
one. Suddenly there are literals of the reference types: these are
values imported from the generator into the generated code.

The problem has been overlooked for more than a decade because none of
the formalizations of staging have been complete enough, and hence do
not handle let-polymorphism along with reference cells and shared CSP.
There is currently no fix for the unsound
let-generalization problem. One solution is proposed in
\cite{kobayashi-poly} but it is restrictive. Another possible solution
is to force the CSP locally-created values to follow the copying
semantics. One may also prohibit generalization if the RHS of a
future-stage let-binding contains an escape, thus introducing the
explicit correlation of staging and let-polymorphism. Along with bad
programs, all these proposals outlaw good ones. Investigating these
trade-offs and finding better ones is the subject of future work. The
present paper does not solve the unsound staged let-generalization
problem either. However, we build a simpler framework to deal with
it, reducing the problem to non-staged generalization.

\section{Translating the Staging Away}
\label{s:translation}

The stymieing problems encountered in the previous section come at the
confluence of staging, let-polymorphism and effects. It is only
natural to wish to investigate them in a simpler setting; for example,
to find a way to translate a staged calculus into the ordinary one.
There have been indeed proposed several `unstaging translations'
\cite{xi-guarded,kameyama-closing,choi-static}, with similar motivations.

Translating the staging away is also practically significant, 
as the method for implementing staged languages. The most attractive is a
source-to-source translation: it lets us implement MetaOCaml just as
a pre-processor to OCaml, fully reusing the existing OCaml compiler
without modifying it (and having to bear the burden of maintaining
the modifications, in sync with the mainline OCaml). This
practical application is the main reason to be interested in unstaging
translations.

Unfortunately, none of the existing unstaging translations deal with
polymorphic let-expressions. Furthermore, an attempt to add them,
described in \S\ref{s:problem}, requires first-class polymorphism,
making the translation unworkable as a source-to-source
translation. Despite its attractiveness, the approach is a dead-end~--
as has been widely acknowledged, including by the author.

We first describe in \S\ref{s:combinators} how well the translation
approach works without the polymorphic let, before illustrating how it
does not with it. \S\ref{s:new-translation} introduces the
solution along with the new questions it poses for let-polymorphism.

\subsection{Staging via Code Combinators}
\label{s:combinators}\mynumberreset

The simplest approach for adding quotation to an existing language is
to write a pre-processor that translates quoted expressions into
ordinary ones, which use pre-defined functions that build and
combine code values, so-called code combinators
\cite{thiemann-combinators,xi-guarded,scala-virtualized}. Code
combinators may of course be used for code generation directly, rather
than through quotation, as has been well demonstrated in Scala
\cite{scala-virtualized}. That said, we will explain code combinators
in the context of an unstaging translation, from the language with
quotations to the language without them~-- motivated by the practical
benefits of such translation.

Our source language, Figure \ref{f:source-target}, is a simple subset
of MetaOCaml (for now, without let-expressions). From now on, we
restrict staging to two-levels only~-- in other words, considering
brackets without nesting~-- as this turns out the overwhelmingly
common use of staged languages. The constants of the language are
integer |i| and string |s| literals and the empty list. Besides
abstraction and application the language includes pairs, 
consing to a list and the creation and dereference of reference cells.
We take the mutation function
|rset: 'a list ref -> 'a -> 'a list| defined in \S\ref{s:poly}
as a primitive. Cross-stage persistent library identifiers such as
|+| are worked out into the syntax. On the other hand, cross-stage
persistence of other identifiers must be explicitly marked with the
|

\begin{figure}[h]
\begin{tabular}{lll}
Constants & \textsf{c ::=} & 
  |i ALT s ALT [] |
\\
Variables & \textsf{x,y,z,f} 
\\
Expressions & \textsf{e ::=} &
 |x ALT c ALT e e ALT fun x -> e ALT e + e ALT (e,e) ALT e :: e ALT ref e ALT !r ALT rset r|
\\
Staged expressions & \textsf{e +::=} &
 |.<e>. ALT .~e ALT 
\\
\end{tabular}
\caption{Source and target languages for the unstaging translation}
\label{f:source-target}
\end{figure}

The target language of the translation is OCaml, without `Meta', i.e.,
without the staged expressions. On the other hand, it has
additional constants for code generation, defined by the following
signature
\begin{code}
module type Code = sig
  type +'a cod

  val int:  int -> int cod
  val str:  string -> string cod
  val add:  int cod -> int cod -> int cod
  val lam:  ('a cod -> 'b cod) -> ('a->'b) cod
  val app:  ('a->'b) cod -> ('a cod -> 'b cod)

  val pair: 'a cod -> 'b cod -> ('a * 'b) cod
  val nil:  'a list cod
  val cons: 'a cod -> 'a list cod -> 'a list cod

  val ref_:  'a cod -> 'a ref cod
  val rget:  'a ref cod -> 'a cod
  val rset:  'a list ref cod -> 'a cod -> 'a list cod

  val csp:   'a -> 'a cod (* CSP local values *)
end
\end{code}
The signature specifies the collection of
typed combinators to generate code for our subset of OCaml:
|int 1| builds the literal |1| code, |add| combines two pieces
of code into the addition expression, etc.  The combinator |lam| builds
the code of a function; its argument is an OCaml function that
returns the code for the body upon receiving the code for the bound
variable. A MetaOCaml expression like
\begin{code}
fun x -> .<fun y -> (y + 1) :: .~x>.
\end{code}
then corresponds to the plain OCaml expression with the code
combinators:
\begin{code}
fun x -> lam (fun y -> cons (add y (int 1)) x)
\end{code}

Formally the unstaging translation is specified in Figure
\ref{f:unstaging}, with two sets of mutually recursive rules: 
|SRB e SRE| deals with the present-stage expressions of the source language
and |TRB e TRE| handles expressions within brackets. The former is
essentially identity, with the single non-trivial rule for brackets.
The translation seems straightforward, which is a great surprise since
the related unstaging translations \cite[\S3]{chen-meta-programming}
and \cite{kameyama-closing,choi-static} are all excruciatingly more
complex and type-directed. The shown translation is novel, which will
become apparent as we discuss the implementation of the |Code| signature
later.

\begin{figure}
Translation at the present-stage |SRB e SRE|
\begin{tabular}[L]{p{0.4\textwidth}p{0.4\textwidth}}
\begin{tabular}{l@{$\quad\mapsto\quad$}l}
|SRB x SRE| & |x| \\
|SRB c SRE| & |c| \\
|SRB e1 e2 SRE| & |SRB e1 SRE SRB e2 SRE| \\
\end{tabular}
&
\begin{tabular}{l@{$\quad\mapsto\quad$}l}
|SRB fun x -> e SRE| & |fun x -> SRB e SRE| \\
\multicolumn{2}{c}{$\ldots$}\\
|SRB .<e>. SRE| & |TRB e TRE| \\
\end{tabular}
\end{tabular}

\bigskip
Translation at the future-stage |TRB e TRE|
\begin{tabular}[L]{p{0.4\textwidth}p{0.4\textwidth}}
\begin{tabular}{l@{$\quad\mapsto\quad$}l}
|TRB x TRE| & |x| \\
|TRB i TRE| & |int i| \\
|TRB s TRE| & |str i| \\
|TRB [] TRE| & |nil| \\
|TRB e1 + e2 TRE| &  |add TRB e1 TRE TRB e2 TRE| \\
|TRB (e1,e2) TRE| &  |pair TRB e1 TRE TRB e2 TRE| \\
|TRB e1 :: e2 TRE| & |cons TRB e1 TRE TRB e2 TRE| \\
\end{tabular}
&
\begin{tabular}{l@{$\quad\mapsto\quad$}l}
|TRB ref e TRE| & |ref_ TRB e TRE| \\
|TRB !e TRE| & |rget TRB e TRE| \\
|TRB rset e TRE| & |rset TRB e TRE| \\
|TRB e1 e2 TRE| & |app TRB e1 TRE TRB e2 TRE| \\
|TRB fun x -> e TRE| & |lam (fun x -> TRB e TRE)| \\
|TRB .~e TRE| & |SRB e SRE| \\
|TRB 
\end{tabular}
\end{tabular}
\caption{Unstaging translation}
\label{f:unstaging}
\end{figure}

\begin{figure}
\begin{tabular}[L]{@{}p{0.23\textwidth}p{0.26\textwidth}p{0.32\textwidth}}
\begin{tabular}{l@{$\;\mapsto\;$}l}
|TRB t code TRE| & |TRB t TRE cod|\\
\multicolumn{2}{c}{identity otherwise}
\end{tabular}
&
\begin{tabular}{l@{$\quad\mapsto\quad$}l}
|TRB x$^0$: t TRE| & |x$^0$: t|
\\
|TRB x$^1$: t TRE| & |x$^0$: t cod|
\end{tabular}
&
\begin{tabular}{Ml@{$\quad\mapsto\quad$}Ml}
 \lceil\Ju{\Gamma}{0}{e}{t}\rceil &
\Ju{\lceil\Gamma\rceil}{0}{\ensuremath\lfloor e\ensuremath\rfloor}
  {\ensuremath\lceil t \ensuremath\rceil}
\\
 \lceil\Ju{\Gamma}{1}{e}{t}\rceil &
\Ju{\lceil\Gamma\rceil}{0}{\ensuremath\lceil e\ensuremath\rceil}%
   {\ensuremath\lceil t \ensuremath\rceil cod}
\end{tabular}
\end{tabular}
\caption{Translation for types, typing environments and judgments}
\label{f:unstaged-typing}
\end{figure}

Our translation is clearly syntax-directed but not
type-directed. Hence it is a source-to-source translation, which can
be done by a macro-processor such as camlp4 or a stand-alone
pre-processor. The rest of the language system (type-checking,
code-generation, standard and user-defined libraries) is used as it
is.

The second property of the translation is that bindings within
brackets are translated to ordinary lambda-bindings.  Coupled with
the appropriate implementation of the |lam| combinator, this property
makes it easy to ensure hygiene.  Correspondingly, variables bound
within brackets are translated to the ordinary, present-stage
variables~-- with the change in type from |t| to |t cod|. One can see that
change from the type of |lam|, and more clearly from Figure
\ref{f:unstaged-typing}, which extends the translation to the typing
judgments and environments described Figure \ref{f:type-system}. The
translation is typing-preserving:
\begin{proposition}
If $\Ju{\Gamma}{n}{e}{t}$ holds then
 $\lceil\Ju{\Gamma}{n}{e}{t}\rceil$ holds as well
\end{proposition}
In other words, a well-typed two-stage MetaOCaml expression is translated into
a well-typed OCaml expression.  The proposition is easily proven by
induction on the typing derivation.  If we also ensure that individual
code combinators produce well-typed code (see below), any typing
errors in the quoted code manifest themselves as OCaml type errors
emitted when type-checking the translated expression. Absent such
errors, the quoted expression, and hence the generated code, are
type-correct.

The following figure shows two implementations of the |Code|
signature. |CodeString| combinators generate ML code as text strings,
justifying their name `code-generating combinators'.  
\begin{code}
module CodeString = struct
  type 'a cod = string

  let int   = string_of_int
  let str x = "\"" ^ String.escaped x ^ "\""
  let add x y = paren @@ x ^ " + " ^ y
  let lam body = 
    let var = gensym "x" in
    "fun " ^ var ^ " -> " ^ body var
  let app f x = paren @@ f ^ " " ^ x
  ...
  let csp x = ... marshaling/unmarshaling ...
end
\end{code}
|CodeReal| is a meta-circular interpreter, representing a code value
as an OCaml thunk (which is also a value).
\begin{code}
module CodeReal = struct
  type 'a cod = unit -> 'a
  open DynBindRef

  let int x = fun () -> x
  let str x = fun () -> x
  let add x y = fun () -> x () + y ()
  let lam body = 
    let r = dnew () in
    let b = body (fun () -> dref r) in
    fun () ->
      let denv = denv_get () in
      fun x -> dlet denv r x b
  let app f x = fun () -> f () (x ())

  let nil = fun () -> []
  ...
  let csp x = fun () -> x
end
\end{code}
The code is utterly trivial, with the exception of |lam|, which does
what a closure has to do: capture the environment at the point
of its creation. We rely on the simple interface for dynamic binding:
\begin{code}
module type DynBind = sig
  type 'a dref
  type denv
  val dnew: unit -> 'a dref
  val dref: 'a dref -> 'a
  val dlet: denv -> 'a dref -> 'a -> (unit -> 'w) -> 'w
  val denv_get: unit -> denv
end
\end{code}
where |dnew| creates a new unbound variable, |dref| dereferences it,
|denv_get| captures the current environment and |dlet denv r x body|
sets the current environment to |denv|, binds |r| to |x| in it and
evaluates the |body|, whose result is returned after the original
environment is restored. The implementation, using either reference
cells or delimited control is straightforward; see the accompanying
source code for details. The source code contains more examples of the
staged translation, including the obligatory factorial: although
our |Code| interface offers neither conditional branching nor
recursive bindings (nor multiplication, for that matter), they are all
obtainable via CSP.

\begin{proposition} If |e : t code| is a program in our subset
of MetaOCaml, then |SRB e SRE : unit->t| is the plain OCaml program
(assuming the |Code| 
interface is implemented by |CodeReal|) such that
|run e| is observationally equivalent
to |SRB e SRE ()|. 
\end{proposition}
Although the intuitions are clear, the rigorous proof of this
proposition is a serious and interesting task. We leave the proof as
a PhD topic.  The proposition justifies the name `unstaging
translation': translating staged OCaml code to plain OCaml. Our
translation is remarkably simple because of the novel implementation
of |lam| in |CodeReal|. The earlier translations had to explicitly
represent and translate the typing and the value environments of an
expression. |DynBind| lets us piggy-back on the typing environment of
OCaml.

One can also intuitively see that |CodeString| and |CodeReal|
correspond: the behavior of the code produced by |CodeString| is the
same as the behavior of running the thunk of |CodeReal| (modulo the
difference in the copying/sharing semantics of CSP). The OCaml
type-checker ensures that any thunk built by |CodeReal| combinators is
well-typed; therefore, it ``will not go wrong'' thanks to the
soundness of OCaml. Hence the code generated by |CodeString| will also
be well-typed and will not go wrong either. The existence of the
|CodeReal| implementation is thus crucial to assuring the soundness of
code generation. Yet another proof of soundness is obtained through
another implementation of |Code|, back into MetaOCaml:
\begin{code}
module CodeCode = struct
  type 'a cod = 'a code

  let int (x:int)    = .<x>.
  let str (x:string) = .<x>.
  let add x y = .<.~x + .~y>.
  let lam body = .<fun x -> .~(body .<x>.)>.
  let app x y = .<.~x  .~y>.
  ...
  let csp x = .<x>.
end
\end{code}
\begin{proposition} If |e : t code| is a program in our subset
of MetaOCaml, then |SRB e SRE : t code| is the equivalent MetaOCaml
program (assuming the |Code| interface is implemented by |CodeCode|):
that is, |e| and |SRB e SRE| have the same side effects and either
both diverge, or return identical (modulo $\alpha$-conversion) code
values.
\end{proposition}
The proof is left as another PhD topic.

Implementing staging by the translation into code
combinators works surprisingly well: Scala's Lightweight Modular
Staging (LMS) is based on similar ideas
\cite{scala-virtualized}. Scheme's implementation of quasi-quote is
also quite alike; only it pays no attention to quoted bindings and
is hence non-hygienic. The translation becomes more complex as we add
to the target language more special forms such as loops, pattern
matching, type annotations, etc.  They pose problems, but they can and
have been dealt with, e.g., in \cite{scala-virtualized}. What could
not be dealt with is let-polymorphism.

\subsection{The Let-Polymorphism Problem}
\label{s:problem}\mynumberreset

The staging translation runs into the roadblock once we add polymorphic
let-bindings, to handle expressions such as those shown in
\S\ref{s:poly-stage}, repeated for reference below.
\begin{code}
.<let x = [] in (2::x,"3"::x)>.
\end{code}
\begin{code}
.<let f = fun x -> x in (f 2, f "3")>.
\end{code}
It may seem we merely need to add to the |Code| signature the combinator
that combines |app| and |lam|:
\begin{code}
val let_ : 'a cod -> ('a cod ->'b cod) -> 'b cod
\end{code}
and the corresponding translation rule
\begin{tabular}[C]{l@{$\quad\mapsto\quad$}l}
|TRB let x = e1 in e2 TRE| &
|let_ TRB e1 TRE (fun x -> TRB e2 TRE)|
\end{tabular}
analogous to |lam|. Then (1) is translated to
\begin{code}
let_ nil (fun x -> pair (cons (int 2) x) (cons (str "3") x))
\end{code}
which, unfortunately, does not type-check.

Recall that our unstaging translation maps bindings in the quoted
code to ordinary lambda-bindings. This exactly is the problem:
unlike let-bindings, lambda-bindings in ML are not
generalizable. First-class polymorphism, if available, does not help
since it requires type annotations, which preclude the
source-to-source translation, done before type checking.

Let-polymorphism hence is the show-stopper for the unstaging
translation. However attractive, we cannot use the translation for
implementing MetaOCaml (unless we give up on polymorphic let within
brackets, which is unpalatable). Therefore, MetaOCaml currently takes the steep
implementation route: modifying the OCaml front-end to account for
brackets and escapes, and the painful patching of the type-checker to
implement the staged type system of Figure \ref{f:type-system}. After
the type-checking, the staging constructs are eliminated by a variant
of the unstaging translation \cite{ber-design}. That translation
manipulates OCaml's |Typedtree|, which represents the AST after
type-checking. Although the tree bears OCaml types, it is `untyped':
it is the ordinary data structure that does not enforce any typing or
scoping invariants. Manipulating the tree is error-prone, with no
(mechanically checked) assurances of correctness.

\section{A New Translation of Quoted \textsf{let}-expressions}
\label{s:new-translation}\mynumberreset

We now present the new translation for quoted let-expressions, which
works even with polymorphic let-bindings. We attempt at the `rational
derivation' of the translation, with our constant refrain of copying
vs. sharing.

The previous \S\ref{s:problem} showed a straightforward translation for
quoted let-expressions, which converts
\begin{code}
.<let x = 1::[] in 
 (2::x,3::x)>
\end{code}
(the quoted version of the first example of \S\ref{s:poly}) to the following
code-combinator based code
\begin{code}
let_ (cons (int 1) nil) @@ fun x ->
pair (cons (int 2) x) (cons (int 3) x)
\end{code}
This example does not have let-polymorphism. But if it did, we are in
trouble: the |x| let-binding of (1) is converted to the |x|
lambda-binding of (2). In the Hindley-Milner type-system lambda-bindings,
unlike let-bindings, are not generalizable. We see the dead end,
regardless of how the |let_| combinator is implemented.

To have any hope of generalization, we need a translation that could map a
let-binding in the quoted code to a let-binding. 
The putative translation should convert (1) into something like
\begin{code}
comb1 (let x = comb2 (cons (int 1) nil) in 
       comb3 (pair (cons (int 2) x) (cons (int 3) x)))
\end{code}
where |comb1|, |comb2|, and |comb3| are yet to be determined
combinators. This proposal seems to be  the most general
compositional, syntax-directed translation that has the desired
let-binding. It fits within the unstaging translation of
\S\ref{s:combinators} in other ways: the future-stage variable |x| in
(1) of the type |int list| is mapped in (3) to the present-stage
variable of the expected (see Figure \ref{f:unstaged-typing}) type
|int list cod|. After all, this the only type that makes, say, 
|cons (int 2) x| well-typed.

All is left is to appropriately implement |comb1|, |comb2| and |comb3|,
for all realizations of the |Code| interface. Proposition 3 imposes a
constraint: Evaluating (3) with the |CodeCode| implementation should
give back (1). And here we notice something odd.  The expression
|(cons (int 1) nil)| evaluates to |.<[1]>.|, according to the existing
code-combinators of |CodeCode|.  The result of |comb2 (cons (int 1) nil)| 
should hence be or contain that singleton list; let us write it
as |.<...[1]...>.|. The let-expression in (3) then produces 
|comb3 .<(2::(...[1]...)),(3::(...[1]...))>.|.  Code-generating combinators
may only combine pieces of code received as arguments but can never
deconstruct or examine them. Therefore, it does not seem possible that
our result can lead to (1), regardless of what |comb1| or |comb3|
might do. We have already inlined |.<[1]>.|, which we should have
let-bound and shared instead.

The only way forward is to have |comb1 .<[1]>.| to somehow generate something
like |.<let y = [1]| |in body>.| and return the let-bound variable as a code
value, that is |.<y>.|. That does not seem possible either. To build
code for a let-expression we need the code for the RHS of the
binding, and the code for the body. The combinator |comb1| does get the
RHS code as the argument; but where is the body?

Fortunately we are stuck at the opportune place: the problem we are
facing is real~-- but it has been solved long time ago in the
partial-evaluation community. The solution is called `let-insertion'
\cite{bondorf-improving,lawall-continuation-based} and requires access
to continuations. The |delimcc| library of OCaml
\cite{caml-shift-tcs} has exactly the control operators needed to
implement the let-insertion interface\footnote{\relax
This let-insertion interface is introduced here for the sake
 of translating quoted expressions and hence the pattern of use
for \textsf{genlet} and \textsf{new\_scope} is determined by 
the translation.}:
\begin{code}
type 'a scope
val new_scope: ('w scope -> 'w cod) -> 'w cod
val genlet: 'w scope -> 'a cod -> 'a cod
\end{code}
These combinators can be used as follows:
\begin{code}
new_scope @@ fun p ->
  lam (fun x-> add x (genlet p (add (int 1) (int 2))))
\end{code}
With the |CodeCode| implementation below it generates 
|.<let y = 1+2 in fun x -> x + y>.|, which \emph{shares} the result of the
subexpression |1+2| across all invocations of the function. In other
words, |genlet p e| inserts, at the place marked by the corresponding
|new_scope|, a |let| statement that binds |e| to a fresh variable, and
returns the code with the name of that variable. We can
finally complete the tentative translation (3): 
\begin{code}
new_scope @@ fun p -> 
  let x = genlet p (cons (int 1) nil) in 
  pair (cons (int 2) x) (cons (int 3) x)
\end{code}
With the |CodeCode| implementation of the combinators that expression
indeed evaluates to (1).

Formally, the new translation of let-expressions takes the form
\begin{code}
TRB let x = e1 in e2 TRE ==>
new_scope (fun p -> let x = genlet p TRB e1 TRE in TRB e2 TRE)
\end{code}
Our running example with let-polymorphism, example (2) from \S\ref{s:poly}
repeated below
\begin{code}
.<let x = [] in (2::x,"3"::x)>.
\end{code}
is hence translated to 
\begin{code}
new_scope @@ fun p ->
let x = genlet p nil in 
pair (cons (int 2) x)
     (cons (str "3") x)
\end{code}
which type-checks, and (with the |CodeCode| combinators) gives back
(8). Incidentally, the combinator code without |genlet|
\begin{code}
new_scope @@ fun p ->
let x = nil in 
pair (cons (int 2) x)
     (cons (str "3") x)
\end{code}
also type-checks. However, it generates
\begin{code}
.<(2::[],"3"::[])>.
\end{code}
where |[]| is inlined rather than shared. The |genlet| combinator hence
implements the sharing in the generated code rather than in the
generator. The fact that the let-variable |x| in (9) gets the polymorphic
type  is the indication, and the vindication, of the equivalence
of copying and sharing in this case. Although the RHS of the
let-binding in (9) is an expression~-- moreover, an effectful
expression, as we are about to see~-- the generalization happens
anyway, thanks to the relaxed value restriction, recalled in \S\ref{s:poly}.
The type variable in |'a list cod| occurs in the covariant position:
note the covariance annotation |+'a cod| in the |Code|
signature.

The code that should not type check in MetaOCaml
\begin{code}
.<let x = ref [] in (rset x 2, rset x "3")>.  (* Does not type-check! *)
\end{code}
is translated to
\begin{code}
new_scope @@ fun p ->
let x = genlet p (ref_ nil) in 
pair (rset x (int 2))
     (rset x (str "3"))
(* Does not type-check! *)
\end{code}
and is rejected by OCaml as expected: the type variable |'a| in the
inferred type |'a list ref cod| for |x| is non-variant and is not
generalized; |x| does not get the polymorphic type and hence cannot be
used in the differently typed contexts.

We implemented |genlet|, directly based on
\cite{lawall-continuation-based}, for all three realizations of the
|Code| signature: not just for |CodeString| but also for |CodeReal| and
|CodeCode|, to demonstrate soundness:
\begin{code}
module CodeLetReal = struct
  include CodeReal
  open Delimcc   type 'a scope = 'a cod prompt
  let new_scope body = let p = new_prompt () in push_prompt p (fun () -> body p)
  let genlet p e = shift0 p (fun k -> let t = e () in k (fun () -> t))
end
\end{code}
\begin{code}
module CodeLetString = struct
  include CodeString
  open Delimcc   type 'a scope = 'a cod prompt
  let new_scope body = ... the same
  let genlet p e = 
   let tvar = gensym "t" in
   shift0 p (fun k -> "let " ^ tvar ^ " = " ^ e ^ " in " ^ k tvar)
end
\end{code}
\begin{code}
module CodeLetCode = struct
  include CodeCode
  open Delimcc   type 'a scope = 'a code prompt
  let new_scope body = ... the same ...
  let genlet p e = shift0 p (fun k -> .<let t = .~e in .~(k .<t>.)>.)
end
\end{code}
(The code of |new_scope| is identical in all three
implementations, although the realizations of the abstract type
|'a scope| differ.)
The type |'a prompt| and the delimited control operators |push_prompt|
and |shift0| are provided by the |delimcc| library
\cite{caml-shift-tcs}.

The |genlet| is so powerful that it easily moves bound variables
\begin{code}
new_scope @@ fun p ->
  lam (fun x-> add x (genlet p (add x (int 2))))
\end{code}
resulting in the generated code 
|let y = x + 2 in fun x -> x + y| with the unbound variable |x|. One
may prevent such undesirable behavior either with a complex type
system (whose glimpse can be caught in
\cite{hygienic-combinators}) or with a dynamic test, as
implemented in MetaOCaml
\cite{ber-design}. In our case, however, |genlet| appears in the code
solely as the result of the translation of a quoted expression.
Fortunately, our translation of
let-expressions puts |new_scope| ``right above'' |genlet|, never
letting them be separated by a |lam| binding. In this case,
delimited control, which underlies |genlet|, is safe (for proofs, see 
\cite{kameyama-shifting-jfp}).

\subsection{Value Restriction at the Whole New Level}
\label{s:new-level-restriction}\mynumberreset

Alas, our new translation stumbles for the common case,
of polymorphic function bindings such as the following:
\begin{code}
.<let f = fun x -> x  in (f 2, f "3")>.
\end{code}
The translation
\begin{code}
new_scope @@ fun p ->
let f = genlet p (lam (fun x -> x)) in 
   pair (app f (int 1)) (app f (str "3"))
(* Does not type-check! *)
\end{code}
is rejected by OCaml: the |genlet| expression has the 
type |('a->'a) cod|, which is not covariant in |'a|. 
Generalizing expressions of
such types is unsound\footnote{However, if the target
  language of code generation has no `dangerous' effects and does not
  need value restriction, we may as well allow generalizing
  expressions of the type \textsf{t cod} regardless of the variance of
  type variables in \textsf{t}.}: 
otherwise, we will have to accept the following
clearly undesirable code~-- the quoted version of our running villain, 
the bad example (10) of \S\ref{s:poly}.
\begin{code}
.<let f = let r = ref [] in
          fun x -> rset r x
  in (f 1, f "3")>.
(* Does not type-check! *)
\end{code}
whose translation
\begin{code}
new_scope @@ fun p1 -> 
  let f = genlet p1 
    (new_scope @@ fun p2 -> 
       let r = genlet p2 (ref_ nil) in
       lam (fun x -> rset r x)) in
  pair (app f (int 1)) (app f (str "3"))
(* Does not type-check! *)
\end{code}
would have type-checked had we allowed generalization for the |genlet p1|
expression. 

The problematic staged code (3) does not type-check according to the
system of Figure \ref{f:type-system} (and in MetaOCaml): the (GenLet)
rule does not apply because the RHS of the let-binding in (3) is not
syntactically a value. Hence we need something like the value
restriction to likewise prevent generalization in (4) while still
allowing it in (2).

Therefore, we amend the translation of let-expressions, 
(7) in \S\ref{s:new-translation}, with the following
\begin{code}
TRB let x = fun z -> e1 in e2 TRE ==>
new_scope (fun p -> let x = genletfun p (fun z -> TRB e1 TRE) in TRB e2 TRE)
\end{code}
where
\begin{code}
val genletfun: 'w scope -> ('a cod -> 'b cod) -> ('a->'b) cod
(* provisional! *)
\end{code}
is a new code-combinator to be added to the let-insertion
interface. In other words, our translation should recognize when a
|let|-bound expression is syntactically a function, and use
|genletfun| rather than the general |genlet| combinator.

With the amended translation, the good example (1) is translated as
\begin{code}
new_scope @@ fun p ->
let f = genletfun p (fun x -> x) in 
pair (app f (int 1))
     (app f (str "3"))
(* See the refined version below! *)
\end{code}
and will type-check. The translation (4) of the bad example (3) will
have to use |genlet| rather than |genletfun| since the RHS of the
let-expression in (3) is not syntactically a function. As we said, (4)
does not actually type-check.

We have thus separated the let-insertion combinators into the general
|genlet| and the specific |genletfun|, which applies only to
the translation of what looks like a function. (We need similar
|genletX| for other polymorphic values of non-covariant types, which
are rare.) For |genlet|, generalization occurs only for covariant
type variables; for |genletfun|, the generalization should occur
always. 

There remains a question how to make the generalization to always
occur for |genletfun| expressions like those in (7), short of
modifying the OCaml compiler. Incidentally, even |Obj.magic| does not
seem to help us with expressions that the relaxed value restriction
cannot generalize: an application of |Obj.magic| is not syntactically
a value. The answer is admittedly a hack;
nevertheless, it gives us another standpoint, however awkward, to hear
the refrain of copying and sharing. And it also works with the extant
OCaml compiler.

Let us step back to look at the clearly flawed translation of (1)
\begin{code}
let f = fun () -> (lam (fun x -> x)) in 
pair (app (f ()) (int 1))
     (app (f ()) (str "3"))
\end{code}
and contemplate what is wrong with it. On the upside, the translated 
expression (8) does type-check: |f| is bound to a thunk (syntactically 
a value) and its type is hence generalized through the ordinary 
value restriction. Since |f| is bound to a thunk we
have to add explicit |()| applications at each place it is used.
Evaluating (8) with the |CodeString| implementation of code
combinators shows the generated code
|((fun x2 -> x2 1), (fun x1 -> x1 "3"))|, with the inlined
rather than shared identity function. 
We had rather the identity function be let-bound and shared. Having
learned that |genlet| introduces let-bindings into the generated code,
the next attempt at the translation of (1) is
\begin{code}
new_scope @@ fun p ->
let f = fun () -> genlet p (lam (fun x -> x)) in 
pair (app (f ()) (int 1))
     (app (f ()) (str "3"))
\end{code}
It also type-checks, since |f| is still bound to a thunk. The
generated code
\begin{code}
let t2 = fun x1 -> x1 in 
let t4 = fun x3 -> x3 in 
((t4 1), (t2 "3"))
\end{code}
is still unsatisfactory: we had rather the two applications in the
pair used the same binding of the identity function.  When |f ()| in
(9) is
first evaluated, it generates a let-binding and returns
the code with the bound variable. We want the second invocation of 
|f ()| to return the code for the very same bound variable. In other
words, we would like to memoize |f|.  Memoization \cite{michie-memo}
indeed was meant to make copying behave like sharing.

The trick hence is introducing a thunk into the let-binding in the
translation to get around the generalization problem and introducing
memoization to restore the sharing destroyed by thunking. In effect,
we do `double memoization': using |genlet| to `memoize' the identity
function in the generated code and memoize the invocation of |genlet|
at the present stage. Once this is understood, the rest is
straightforward. To make the translation similar to (7), we combine
|genlet| with the memoization into |genletfun|:
\begin{code}
type 'w funscope
val new_funscope : ('w funscope -> 'w cod) -> 'w cod
val genletfun: 'w funscope -> ('a cod -> 'b cod) -> ('a->'b) cod
\end{code}
The final translation of (1) then reads:
\begin{code}
new_funscope @@ fun p ->
let f = fun () -> genletfun p (fun x -> x) in 
pair (app (f ()) (int 1))
     (app (f ()) (str "3"))
\end{code}
Unlike (7), we had to replace the occurrence of |f| with |f ()|~--
explicitly marking the type instantiation, so to speak. This complication is
still possible to implement with the source-to-source translation
(call-by-name let-binding of \cite{Leroy-cbn-poly} would be really
handy here).

The double-memoizing |genletfun| can be easily and generically implemented,
with a small bit of magic
\begin{code}
type afun = | AFun : ('a -> 'b) cod -> afun
            | ANone : afun
type 'w funscope = 'w scope * afun ref
let new_funscope body = new_scope (fun p -> body (p, ref ANone))
let genletfun : 'w funscope -> ('a cod -> 'b cod) -> ('a->'b) cod =
  fun (p,r) body ->
  match !r with
  | ANone -> let fn = lam body in
             let x = genlet p fn in
             r := AFun x; x
  | AFun x -> Obj.magic x
\end{code}
The code uncannily resembles (4) of
\S\ref{s:poly-stage}.
 
One may wonder if it would be better to add |genletfun| to the OCaml
type-checker as an ad hoc, always-generalize case. The answer at
present should be ``no'': |genletfun| is still unsound, in the edge
case of (4) of \S\ref{s:poly-stage}~-- the example that is also
unsound in the present MetaOCaml. Here is this example again, for
reference
\begin{code}
.<let f = fun () -> .~(lift (ref [])) in 
     (rset (f ()) 2, rset (f ()) "3")>.
\end{code}
Its translation
\begin{code}
new_funscope @@ fun p ->
 let f = fun () -> genletfun p (fun _ -> csp (ref [])) in
 pair (rset (app (f ()) (csp ())) (int 1))
      (rset (app (f ()) (csp ())) (str "3"))
\end{code}
type-checks~-- and when run with |CodeReal| exhibits the same
segmentation fault it does in the case of the corresponding MetaOCaml
code. 

It seems our unstaging translation is just as sound~-- or
unsound~-- as MetaOCaml. Solving the soundness problem of MetaOCaml described
in \S\ref{s:poly-stage} will make, we conjecture, the unstaging
translation fully sound as well. Much work lies ahead.

\section{Conclusions}

We have presented a new, typing-preserving translation from a
higher-order typed staged language, with hygienic quotations and
unquotations, to the language without quotations. Code-generation is
accomplished through a library of code-generation combinators.  Our
translation is remarkably simpler than other unstaging translations: it is
not type-directed and can be accomplished as a source-to-source
transformation. Mainly, the translation works for polymorphic let:
let-expressions
within quotes are transformed to also let-expressions, hence preserving
generalization. All throughout the presentation we emphasized deep
connections, between polymorphism and sharing.

Our translation is already a viable method of
implementing staged languages. Yet the theoretical work has
just began. Yet another feature of our translation is
`bug-preservation': the restrictions and unsound edge cases of
let-polymorphic expressions are preserved in the translation. The
problems hence can be investigated in a simpler setting, without
staging.

We thus propose a research program:
\begin{enumerate}
\item 
Formally establishing the equivalence properties of |CodeReal|,
|CodeCode| and |CodeString| and formally justifying the
translation;
\item Generalizing  from two-stage to multiple-stages,
  that is, to multiple levels of quotations;
\item Proving that the edge case described in
\S\ref{s:poly-stage} is the only one where |genletfun| is unsound;
\item Relaxing the value restriction even more so that |genletfun|
  could be implemented without magic;
\item Investigating trade-offs of various solutions to the unsoundness
  problem in \S\ref{s:poly-stage} and finding the solution with the
  least loss in expressiveness and convenience.
\end{enumerate}

\subsubsection*{Acknowledgments}
I thank Jacques Garrigue and Atsushi Igarashi for many helpful
discussions. Comments and suggestions by Yukiyoshi Kameyama and
anonymous reviewers are gratefully appreciated. This work was
partially supported by JSPS KAKENHI Grant Numbers 22300005, 25540001,
15H02681.

\bibliographystyle{eptcs}
\bibliography{../metafx}
\end{document}